\documentclass[12pt,aps,prd,preprint,tightenlines,superscriptaddress,
  showpacs,nofootinbib]{revtex4}
\newcommand{\PRE}[1]{{#1}} 

\usepackage{hyperref}      
\usepackage{bm}
\usepackage{epsfig}
\usepackage{ color}

\newcommand{\gweak}{g_{\text{weak}}}
\newcommand{\mweak}{m_{\text{weak}}}

\newcommand{\sigmaan}{\sigma_{\text{an}}}

\newcommand{\ev}{\text{eV}}

\newcommand{\mev}{\text{MeV}}
\newcommand{\gev}{\text{GeV}}
\newcommand{\tev}{\text{TeV}}
\newcommand{\pb}{\text{pb}}

\newcommand{\cm}{\text{cm}}

\newcommand{\km}{\text{km}}

\newcommand{\s}{\text{s}}

\newcommand{\yr}{\text{yr}}

\newcommand{\kpc}{\text{kpc}}

\newcommand{\ie}{{\em i.e.}}

\newcommand{\eqref}[1]{Eq.~(\ref{#1})}

\newcommand{\secref}[1]{Sec.~\ref{sec:#1}}

\newcommand{\figref}[1]{Fig.~\ref{fig:#1}}

\newcommand{\mgravitino}{M_{3/2}}

\newcommand{\photino}{\tilde{\gamma}}
\newcommand{\mphotino}{m_{\tilde{\gamma}}}

\newcommand{\Omegatot}{\Omega_{\text{tot}}}
\newcommand{\neff}{N_{\text{eff}}}

\newcommand{\vev}[1]{\langle #1\rangle}

\begin{document}

\preprint{UCI-TR-2011-16}

\title{ \PRE{\vspace*{1.5in}} 
WIMPless Dark Matter in Anomaly-Mediated Supersymmetry Breaking with
Hidden QED
\PRE{\vspace*{0.3in}} }

\author{Jonathan L.~Feng}
\affiliation{Department of Physics and Astronomy, University of
California, Irvine, CA 92697, USA
\PRE{\vspace*{.2in}}
}

\author{Vikram Rentala}
\affiliation{Department of Physics, University of Arizona, Tucson, AZ
  85721, USA
\PRE{\vspace*{.2in}}
}
\affiliation{Department of Physics and Astronomy, University of
California, Irvine, CA 92697, USA
\PRE{\vspace*{.2in}}
}

\author{Ze'ev Surujon\PRE{\vspace*{.5in}}}
\affiliation{Department of Physics and Astronomy, University of
California, Irvine, CA 92697, USA
\PRE{\vspace*{.2in}}
}
\affiliation{Department of Physics and Astronomy, University of
California, Riverside, CA 92521, USA
\PRE{\vspace*{.5in}}
}

\date{August 2011}

\begin{abstract}
\PRE{\vspace*{.3in}} In anomaly-mediated supersymmetry breaking,
superpartners in a hidden sector have masses that are proportional to
couplings squared, and so naturally freeze out with the desired dark
matter relic density for a large range of masses.  We present an
extremely simple realization of this possibility, with WIMPless dark
matter arising from a hidden sector that is supersymmetric QED with
$N_F$ flavors.  Dark matter is multi-component, composed of hidden
leptons and sleptons with masses anywhere from 10 GeV to 10 TeV, and
hidden photons provide the thermal bath.  The dark matter
self-interacts through hidden sector Coulomb scatterings that are
potentially observable.  In addition, the hidden photon contribution
to the number of relativistic degrees of freedom is in the range
$\Delta \neff \sim 0 - 2$, and, if the hidden and visible sectors were
initially in thermal contact, the model predicts $\Delta \neff \sim
0.2 - 0.4$.  Data already taken by Planck may provide evidence of such
deviations.
\end{abstract}


\maketitle

\section{Introduction}
\label{sec:introduction}

The thermal relic density of a particle $X$ has the parametric
dependence
\begin{equation}
\Omega_X \propto \frac{1}{\langle \sigmaan v \rangle} \sim
\frac{m_X^2}{g_X^4} \ ,
\label{relicdensity}
\end{equation}
where $\langle \sigmaan v \rangle$ is the thermally-averaged product
of the annihilation cross section and relative velocity, and $m_X$ and
$g_X$ are the characteristic mass scale and coupling determining this
cross section.  The ``WIMP miracle'' is the fact that, for particles
with weak scale masses and interactions, $g_X \sim \gweak \sim 0.6$
and $m_X \sim \mweak \sim 100~\gev - 1~\tev$, the resulting thermal
relic density is $\Omega_X \sim 0.1$, as desired for dark matter.
This coincidence is the primary linchpin connecting the high energy
and cosmic frontiers, and it has driven much of the research, both
theoretical and experimental, in dark matter for decades.

At the same time, as \eqref{relicdensity} makes clear, the required
thermal relic density may be realized by lighter and more weakly
interacting particles (or heavier and more strongly interacting
particles), provided their masses and couplings scale together to
leave $\Omega_X$ invariant.  Remarkably, particles with exactly these
properties are found in supersymmetric theories that are motivated not
by cosmology, but by particle physics~\cite{Feng:2008ya,Feng:2008mu}.
This fact, the ``WIMPless miracle,'' suggests that, at least in the
context of supersymmetry, the requirement of a correct thermal relic
density does not necessarily lead to WIMPs and all of their
accompanying phenomena, but rather argues for a more general viewpoint
with a broader set of implications for both experiments and
astrophysical observations~\cite{Feng:2008dz,Feng:2008qn,%
Kumar:2009ws,Barger:2010ng,Zhu:2011dz,McKeen:2009rm,Yeghiyan:2009xc,%
McKeen:2009ny,Badin:2010uh,Alwall:2010jc,Goodman:2010ku,Alwall:2011zm}.

In this work, we consider supersymmetric models with anomaly-mediated
supersymmetry breaking (AMSB)~\cite{Randall:1998uk,Giudice:1998xp} and
a hidden sector, that is, a set of particles with no standard model
gauge interactions.  In AMSB, the standard model is sequestered from
the supersymmetry-breaking sector, for example, by being located on a
different brane in extra dimensions~\cite{Randall:1998uk}, or because
the supersymmetry-breaking sector is near-conformal over some energy
range~\cite{Luty:2001zv}.  This sequestering implies that new physics
contributions to flavor- and CP-violating observables are suppressed,
which is the prime phenomenological virtue of AMSB.\footnote{Both
left- and right-handed sleptons are tachyonic in the minimal
realization of AMSB. Our main concern here will be model-building in
the hidden sector, and we assume that this visible sector problem is
resolved by one of the many proposed mechanisms; see, for example,
Refs.~\cite{Pomarol:1999ie,Chacko:1999am,Katz:1999uw}.}  We assume
that the hidden sector is also sequestered.  The superpartner masses
in both the visible and hidden sectors then have the form
\begin{equation}
\tilde{m} \sim \frac{g^2}{16\pi^2} \mgravitino \ ,
\end{equation}
where $g$ represents the superpartner's interaction couplings and
$\mgravitino \sim 100~\tev$ is the gravitino mass.  Superpartners $X$
in the hidden sector therefore satisfy
\begin{equation}
\frac{m_X}{g_X^2} \sim \frac{1}{16 \pi^2} \mgravitino
\sim \frac{\mweak}{\gweak^2} \ ,
\label{miracle}
\end{equation}
where $m_X$ and $g_X$ are the hidden superpartner's mass and gauge
coupling, and so realize the WIMPless miracle.

Although \eqref{miracle} goes a long way toward providing an
attractive dark matter scenario, it is, of course, not sufficient.  To
realize the WIMPless scenario fully, the dark matter candidate must
also (1) be stable on cosmological time scales and (2) annihilate to
light particles, \ie, the thermal bath, with cross section $\sigmaan
\sim g_X^4/m_X^2$.  In addition, the model must also satisfy basic
constraints, such as vacuum stability and perturbativity, and the dark
matter's properties must be consistent with all experimental and
observational constraints.  These are not trivial constraints in the
context of AMSB models, in which all superpartner masses are
determined by a small number of low-energy parameters.  Of course,
this property also makes the scenarios more predictive, a highly
laudable feature given the usual standards of hidden sector scenarios.

The possibility of WIMPless dark matter in AMSB was previously
considered in Ref.~\cite{Feng:2011ik}.  There, non-Abelian gauge
interactions in the hidden sector were considered, and a number of
viable models were constructed.  Here we consider hidden sectors with
Abelian gauge interactions.  We find that a simple scenario with
hidden supersymmetric QED (SQED) satisfies all constraints, and at the
same time predicts new phenomena that may be tested by near future
astrophysical observations.

\section{SQED in AMSB}
\label{sec:sqed}

\subsection{First Pass: $N_F = 1$}

We consider SQED in the hidden sector, a U(1) gauge interaction with
gauge coupling $g$.  For pure SQED in AMSB, the photino is massless
and interaction-less, and there is no cold dark matter candidate.  We
therefore begin by introducing one flavor of matter, that is, a pair
of superfields $\hat{e}_+$ and $\hat{e}_-$ with charges $+1$ and $-1$,
respectively, which contain component fermions (electrons) $e_{\pm}$
and scalars (selectrons) $\tilde{e}_{\pm}$.

The AMSB soft masses are
well-known~\cite{Randall:1998uk,Giudice:1998xp}, and the relevant
formulae are summarized in the first Appendix of
Ref.~\cite{Feng:2011ik}.  For this simple model, the soft
contributions to the photino mass and selectron masses are
\begin{eqnarray}
\mphotino &=& b \frac{g^2}{16\pi^2} \mgravitino \\
m_{\tilde{e}_{\pm}}^2 &=& - 2 b \left[ \frac{g^2}{16 \pi^2} \mgravitino
\right]^2 \ ,
\end{eqnarray}
where $b=2$ is the one-loop $\beta$-function coefficient.  All
$A$-terms vanish, since there are no Yukawa couplings.  Because the
gauge interaction is not asymptotically-free ($b>0$), the selectrons
are tachyonic.  This is the source of the well-known tachyonic slepton
problem in the visible sector of minimal AMSB models.  The potential
is therefore unstable in the $D$-flat direction $\vev{\tilde{e}_+} =
\vev{\tilde{e}_-}$.  Supergravity interactions could stabilize the
potential, but generically, one would then expect the scalars to
acquire Planck-scale vacuum expectation values, which is incompatible
with the possibility of a WIMPless dark matter candidate.

To stabilize the potential, we may introduce a mass term in the
superpotential $W = \mu \hat{e}_+ \hat{e}_-$.  We do not specify the
source of this mass term; presumably it is generated by a mechanism
similar to one that generates the $\mu$-term in the visible sector.
For now we assume there is a mechanism to generate a mass $\mu \sim
(g^2/16 \pi^2) \mgravitino$, and analyze the phenomenological
implications with $\mu$ as a free parameter.  We return to this
discussion in \secref{discussion} when the phenomenologically-allowed
regions of parameter space have been established.

As an illustration of the utility of this approach, we note that with
this mass term, the physical masses of the particles are
\begin{eqnarray}
\mphotino &=& b \frac{g^2}{16\pi^2} \mgravitino \\
m_e &=& \left| \mu \right| \\
m_{\tilde{e}_{\pm}} &=& \left[ \left| \mu \right| ^2 -
2 b \left( \frac{g^2}{16 \pi^2} \mgravitino \right)^2
\right]^{1/2} \\
m_{\gamma} &=& 0 \ .
\end{eqnarray}
Vacuum stability requires that the selectrons be non-tachyonic.  As a
result, U(1) and $R$-parity are conserved.  Photons are massless and
form the thermal bath, selectrons are stable by U(1) charge
conservation, and electrons are stable by U(1) charge and $R$-parity
conservation.

Unfortunately, vacuum stability also implies $m_e > \mphotino$.  The
decay $\photino \to \tilde{e} e$ is therefore forbidden, and photinos
are also stable.  This is problematic: photinos annihilate through
$\photino \photino \to \gamma \gamma$, which is loop-suppressed, with
cross section $\sigmaan \sim (g^2/16\pi^2)^2 g^4/\mphotino^2$, and so
the photino is overproduced.  This problem persists even if a soft
supersymmetry-breaking $B$-term is introduced.  Thus, for all values
of $\mu$, irrespective of how it is generated, we see that this
scenario cannot provide a viable WIMPless dark matter model.

\subsection{Viable Models: $N_F \ge 2$ }
\label{sec:viable}

We may build a viable model, however, by introducing additional
charged matter fields to raise the photino mass and allow it to decay.
In general, consider $N_F$ pairs of superfields $\hat{e}_{i+}$ and
$\hat{e}_{i-}$, containing component field fermions (leptons)
$e_{i\pm}$ and scalars (sleptons) $\tilde{e}_{i \pm}$, $i = 1, \ldots,
N_F$, with charges $\pm q_i$, supersymmetric masses $\mu_i$, and
bilinear scalar couplings $B_i \tilde{e}_{i+} \tilde{e}_{i-}$.  The
superpotential is
\begin{equation}
W = \sum_{i=1}^{N_F} \mu_i \hat{e}_{i+} \hat{e}_{i-} \ ,
\end{equation}
and the physical masses of the particles in the theory are
\begin{eqnarray}
\mphotino &=& b \frac{g^2}{16\pi^2} \mgravitino \\
m_{e_i} &=& \left| \mu_i \right| \\
m_{\tilde{e}_{i \, 1,2}}^2 &=& \left[ \left| \mu_i \right| ^2 -
2 b q_i^2 \left( \frac{g^2}{16 \pi^2} \mgravitino \right)^2 \pm B_i
\right]^{1/2} \\
m_\gamma &=& 0 \ ,
\end{eqnarray}
where $b = 2 \sum q_i^2$.  This model has several symmetries: the U(1)
gauge symmetry of SQED, a global U(1)$^{N_F}$ lepton flavor symmetry,
and $R$-parity.  As a result of these symmetries, the $N_F$ flavors of
leptons and $N_F$ flavors of sleptons are all stable, naturally
yielding multi-component dark matter with the total relic density
roughly evenly divided between each of the components.

For simplicity, consider a version of the general model with $q_i =
1$, $\mu_i=\mu$, and $B_i=0$ for all $i$.  The lepton flavor symmetry
is enhanced to SU($N_F$).  Relaxing the assumption of universal masses
and charges does not change our main conclusions qualitatively.  As a
very rough acknowledgment of this more general possibility, we present
results for a continuous range of $N_F$, where non-integer $N_F$ may
capture some of the features of these less minimal scenarios.  We also
assume negligible mixing between the visible and hidden photons, as
would be the case if the visible and hidden sectors are sequestered
from each other, just as they are sequestered from the
supersymmetry-breaking sector.  We briefly discuss the possibility of
connectors mediating interactions between the two sectors to
\secref{discussion}.

\begin{figure}[tb]
\includegraphics[width=0.49\columnwidth]{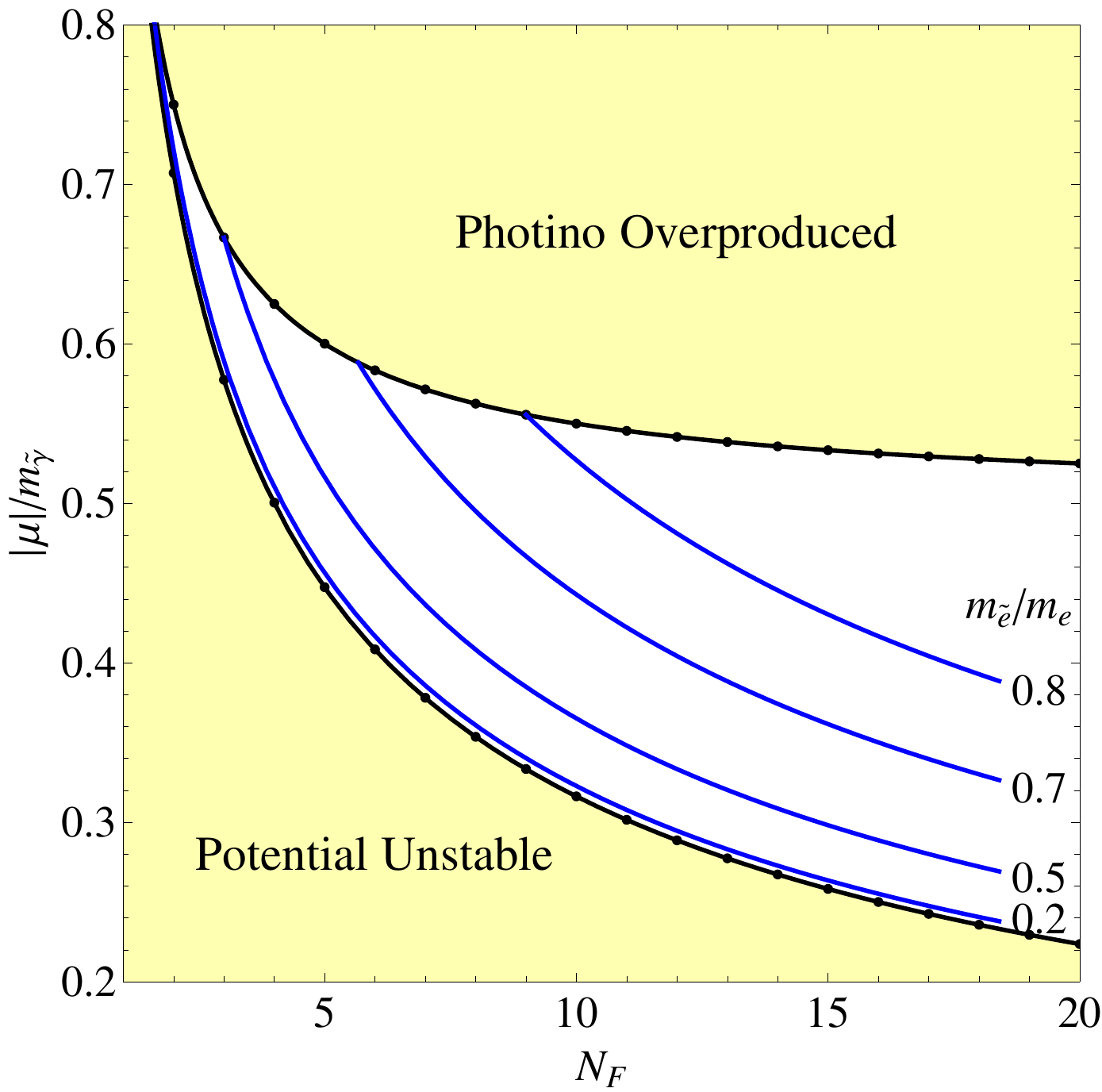}
\vspace*{-.1in}
\caption{Allowed region in the $(N_F, |\mu|/\mphotino)$ parameter
space of WIMPless dark matter from hidden SQED with $N_F$ flavors and
global SU($N_F$) flavor symmetry.  The lower and upper shaded regions
are excluded by the requirements that the vacuum be stable and
photinos decay, respectively.  In the allowed region, multi-component
dark matter is composed of $N_F$ degenerate flavors of hidden leptons
$e_i$ and $N_F$ degenerate flavors of sleptons $\tilde{e}_i$.
Contours of constant $m_{\tilde{e}_i}/m_{e_i} $ are shown.
\label{fig:massratio}}
\end{figure}

The resulting $\beta$-function coefficient is $b=2 N_F$, and the
physical masses are
\begin{eqnarray}
\mphotino &=& 2 N_F \frac{g^2}{16\pi^2} \mgravitino \\
m_{e_i} &=& \left| \mu \right| \\
m_{\tilde{e}_{i \pm}}^2 &=& \sqrt{ \left| \mu \right|^2
- \frac{\mphotino^2}{N_F}} \\
m_\gamma &=& 0 \ .
\end{eqnarray}
The requirements of vacuum stability and photino instability constrain
$|\mu| / \mphotino$ from below and above, respectively.  The resulting
allowed range
\begin{equation}
   \frac{1}{\sqrt{N_F}}<\frac{ | \mu |}{\mphotino}
   <\frac{1}{2}+\frac{1}{2N_F} \ ,
\end{equation}
is shown in \figref{massratio}.  As expected, there is no allowed
range for $N_F = 1$, but for $N_F=2$, there are viable models with
\begin{equation}
   0.71 \alt \frac{|\mu|}{\mphotino} < 0.75 \ ,
\end{equation}
and the allowed range expands as $N_F$ increases. Non-zero $B$-terms
would move both the lower and upper boundaries to larger values of
$|\mu| / \mphotino$. In the allowed region, $m_{\tilde{e}_i}$ and
$m_{e_i}$ are of the same order of magnitude in almost all of the
parameter space, with $m_{\tilde{e}_i} \ll m_{e_i}$ only in a thin
skin layer near the vacuum instability boundary, where
$m_{\tilde{e}_i}$ vanishes.

\section{Relic Density}
\label{sec:relicdensity}

The freeze out of hidden leptons and sleptons may be analyzed in the
usual way, with the slight complication that the hidden sector may be
at a different temperature than the visible sector.  This possibility
has been analyzed in detail~\cite{Feng:2008mu,Das:2010ts}, and the
results are summarized in the second Appendix of
Ref.~\cite{Feng:2011ik}. For a particle $X$ that annihilates through
$S$-wave processes with annihilation cross section
\begin{equation}
\sigmaan v \approx
\sigma_0  \equiv k_X \frac{\pi \alpha_X^2}{m_X^2} \ ,
\end{equation}
where $k_X$ is an ${\cal O}(1)$ constant determined by the specific
annihilation processes, the thermal relic density is
\begin{equation}
\Omega_X \approx \xi_f \, \frac{0.17~\pb}{\sigma_0}
\simeq 0.23 \ \xi_f \, \frac{1}{k_X}
\left[ \frac{0.025}{\alpha_X} \frac{m_X}{\tev} \right]^2 \ ,
\end{equation}
where $\xi_f \equiv T_f^{\text{h}}/T_f^{\text{v}}$ is the ratio of the
hidden to visible sector temperatures when the hidden dark matter
freezes out.

In the present case, both hidden leptons and sleptons annihilate to
hidden photons through $S$-wave processes, and so the results above
apply.  Lepton pair annihilation $e_i \bar{e}_i \to \gamma \gamma$ is
mediated by diagrams with leptons in the $t$- and $u$-channel, as in
standard QED.  Slepton pair annihilation $\tilde{e}_{i\, \pm}
\tilde{e}^*_{i\, \pm} \to \gamma \gamma$ is mediated by similar
diagrams with sleptons in the $t$- and $u$-channel, and also through
four-point interactions.  The annihilation coefficients are $k_{e_i} =
1$ and $k_{\tilde{e}_i} = 2$~\cite{Feng:2008mu,Feng:2009mn}.  The
resulting total relic density is therefore
\begin{eqnarray}
\Omegatot &\simeq& 0.23 \ \xi_f \, N_F \left[
\frac{1}{k_{e_i}} \left(\frac{0.025}{\alpha}
\frac{m_{e_i}}{\tev}\right)^2
+ \frac{2}{k_{\tilde{e}_i}} \left(\frac{0.025}{\alpha}
\frac{m_{\tilde{e}_i}}{\tev}\right)^2 \right] \\
&=& 0.23 \ \frac{N_F^3}{6.3}
\left(2 \frac{|\mu|^2}{\mphotino^2} - \frac{1}{N_F} \right)
\left( \frac{ \sqrt{\xi_f} \, \mgravitino}{100~\tev} \right)^2 \ ,
\end{eqnarray}
where, in the second line, we have substituted the appropriate values
of $k_{e_i, \tilde{e}_i}$ and $m_{e_i, \tilde{e}_i}$.  As required for
a realization of the WIMPless miracle, the relic density is
independent of the gauge coupling, provided $|\mu| \sim \mphotino$, as
discussed above.

\begin{figure}[tb]
\includegraphics[width=0.49\columnwidth]{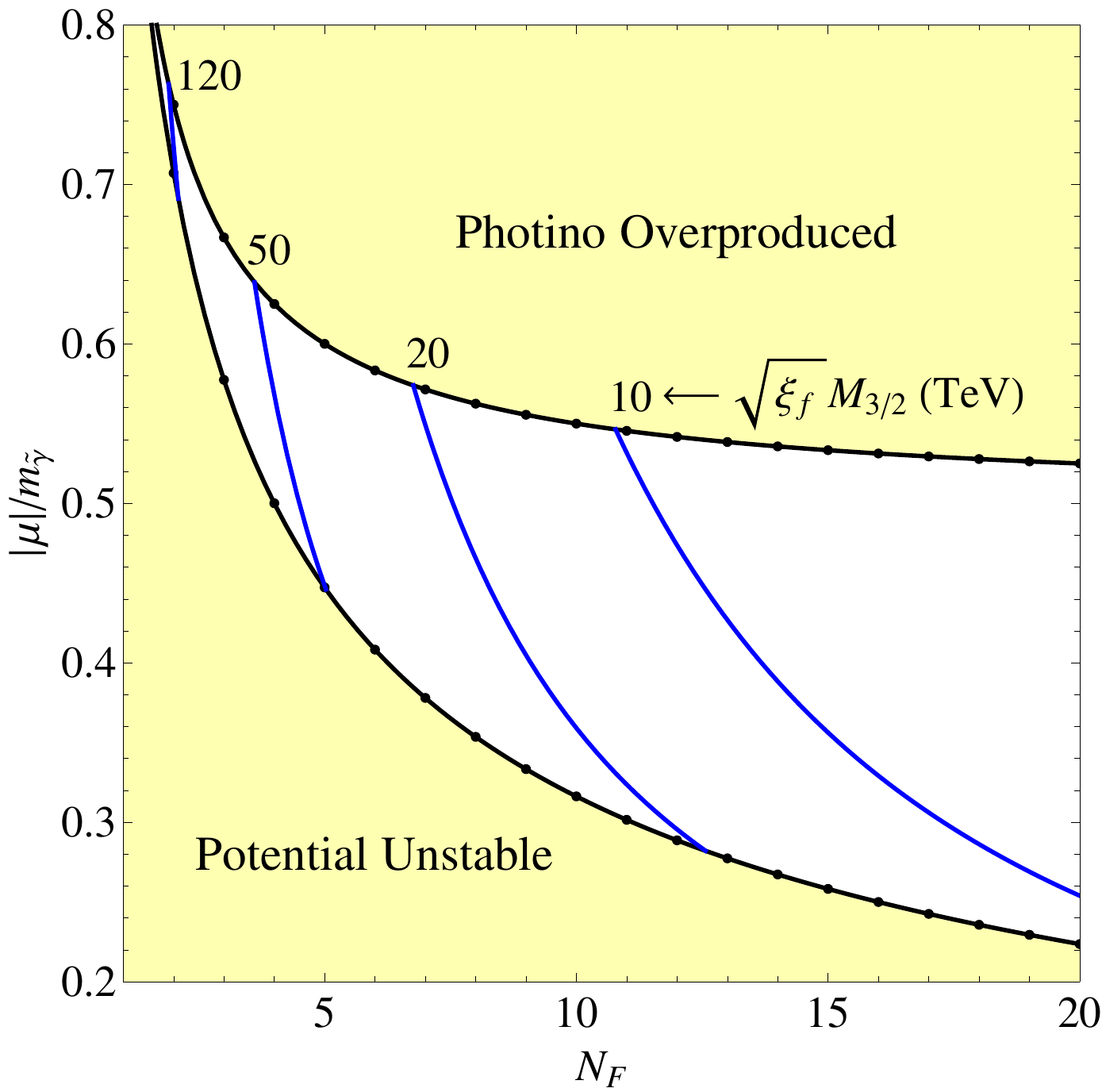}
\vspace*{-.1in}
\caption{Contours of constant $\sqrt{\xi_f} \mgravitino$, as
determined by requiring $\Omegatot \simeq 0.23$ in the $(N_F,
|\mu|/\mphotino)$ parameter space.
\label{fig:mgravitino}}
\end{figure}

In \figref{mgravitino}, we set $\Omegatot \simeq 0.23$ and plot
contours of constant $\sqrt{\xi_f} \mgravitino$ in the $(N_F,
|\mu|/\mphotino)$ plane.  LEP2 constraints require Wino masses
$m_{\tilde{W}} > 92 - 103~\gev$, depending on the chargino-neutralino
mass difference~\cite{Feng:2009te}. Assuming the minimal AMSB relation
for the Wino mass, this implies $\mgravitino \simeq 372 \,
m_{\tilde{W}} \agt 34~\tev$.  The LHC also bounds the AMSB scenario,
but these constraints depend on the spectrum of strongly-interacting
superpartners.  As a conservative example, current bounds from the LHC
require $m_{\tilde{g}} \agt 800~\gev$ for large squark
masses~\cite{LHCbounds}, which implies $\mgravitino \simeq 35 \,
m_{\tilde{g}} \agt 28~\tev$.  These LHC bounds will certainly improve
rapidly.  At the moment, however, the LEP2 bounds may serve as a guide
to what is allowed, and we consider $\mgravitino \approx 34~\tev$ to
be the lower bound, and values of $\mgravitino$ just above 34 TeV to
be the most natural from the point of the gauge hierarchy problem.
For $\xi_f$, perhaps the most motivated values are $\xi_f \sim 1$, as
would result, for example, if the hidden and visible sectors were
similarly reheated through inflaton
decay~\cite{Hodges:1993yb,Berezhiani:1995am}, or, as discussed below,
if they were in thermal contact in the very early Universe.  For these
values of $\mgravitino$ and $\xi_f$, we see from \figref{mgravitino}
that the hidden leptons and sleptons are excellent cold dark matter
candidates, independent of their mass (or equivalently, the gauge
coupling).

For comparison, note that the visible sector's Wino may also be cold
dark matter.  However, given that SU(2) is ``accidentally'' nearly
conformal in the minimal supersymmetric standard model (MSSM), AMSB
Winos annihilate very efficiently and have the correct thermal relic
density only for very large masses $m_{\tilde{W}} \sim
3~\tev$~\cite{Giudice:1998xp}.  In minimal models, this implies
$m_{\tilde{g}} \sim 30~\tev$ and $\mgravitino \sim 1100~\tev$, highly
unnatural values if AMSB supersymmetry is to be a solution to the
gauge hierarchy problem.  The hidden sector scenario analyzed here
rectifies this situation, in the sense that for the natural range
$\mgravitino \sim 100~\tev$ where the relic abundance of visible
sector superpartners is generically too low to contribute
significantly to dark matter, the hidden leptons and sleptons have the
desired abundances to be dark matter.

Finally, perhaps the most motivated possibility for relating the
visible and hidden sector temperatures is to assume that these sectors
were in thermal contact at very early times, but then decoupled.  In
this case, the ratio of temperatures at very early times is
$\xi_{\infty} = 1$, where the subscript ``$\infty$'' denotes the
period of thermal coupling, and the ratio of temperatures at later
times is determined by the visible and hidden sector particle spectra,
assuming that after decoupling the entropy of each sector is
separately conserved.

In particular, the ratio of temperatures at freeze out is related to
$\xi_{\infty}$ by
\begin{equation}
\xi_f = \left[ \frac{g_*^{\text{h}}(T_{\infty}^{\text{h}})}
{g_*^{\text{h}}(T_f^{\text{h}})} \,
\frac{g_*^{\text{v}}(T_f^{\text{v}})}
{g_*^{\text{v}}(T_{\infty}^{\text{v}})} \right]^{\frac{1}{3}}
\xi_{\infty} \ ,
\label{xi}
\end{equation}
where $g_*(T)$ is the number of relativistic degrees of freedom at
temperature $T$.  This depends on the presence or absence of very
heavy degrees of freedom in the visible and hidden sectors.  However,
assuming a desert in the MSSM and our hidden SQED sectors, that is, no
particles with masses between the temperature at which the two sectors
thermally decoupled and the masses of the heaviest particles we have
considered, $g_*^{\text{v}} (T_{\infty}^{\text{v}}) =
g_*^{\text{MSSM}} = 228.75$, the total number of degrees of freedom in
the MSSM, and $g_*^{\text{h}} (T_{\infty}^{\text{h}}) = 7.5 N_F +4$.
At freeze out, we may take $g_*^{\text{h}}(T_f^{\text{h}}) = 2$, since
the temperature at freeze out $T_f^{\text{h}} \sim m_X / 25$ is
generically below all dark matter masses.  In the visible sector,
possible benchmark values for $g_*^{\text{v}}(T_f^{\text{v}})$ are
86.25, its SM value for temperatures between $m_b$ and $m_W$,
$g_*^{\text{SM}} = 106.75$, the total $g_*$ for the standard model,
and $g_*^{\text{MSSM}}=228.75$.  Normalizing to
$g_*^{\text{v}}(T_f^{\text{v}}) = g_*^{\text{SM}}$, we find
\begin{equation}
\xi_f = 1.21 \left[ \left( N_F + \frac{8}{15} \right)
\left( \frac{g_*^{\text{v}}(T_f^{\text{v}})}{106.75}
\right) \right]^{\frac{1}{3}} \xi_{\infty} \ .
\label{xif}
\end{equation}
The ratio of temperatures $\xi$ typically grows between very early
times and freeze out, because almost the entire hidden sector becomes
non-relativistic before freeze out, while this is not true for the
visible sector.

\begin{figure}[tb]
\includegraphics[width=0.49\columnwidth]{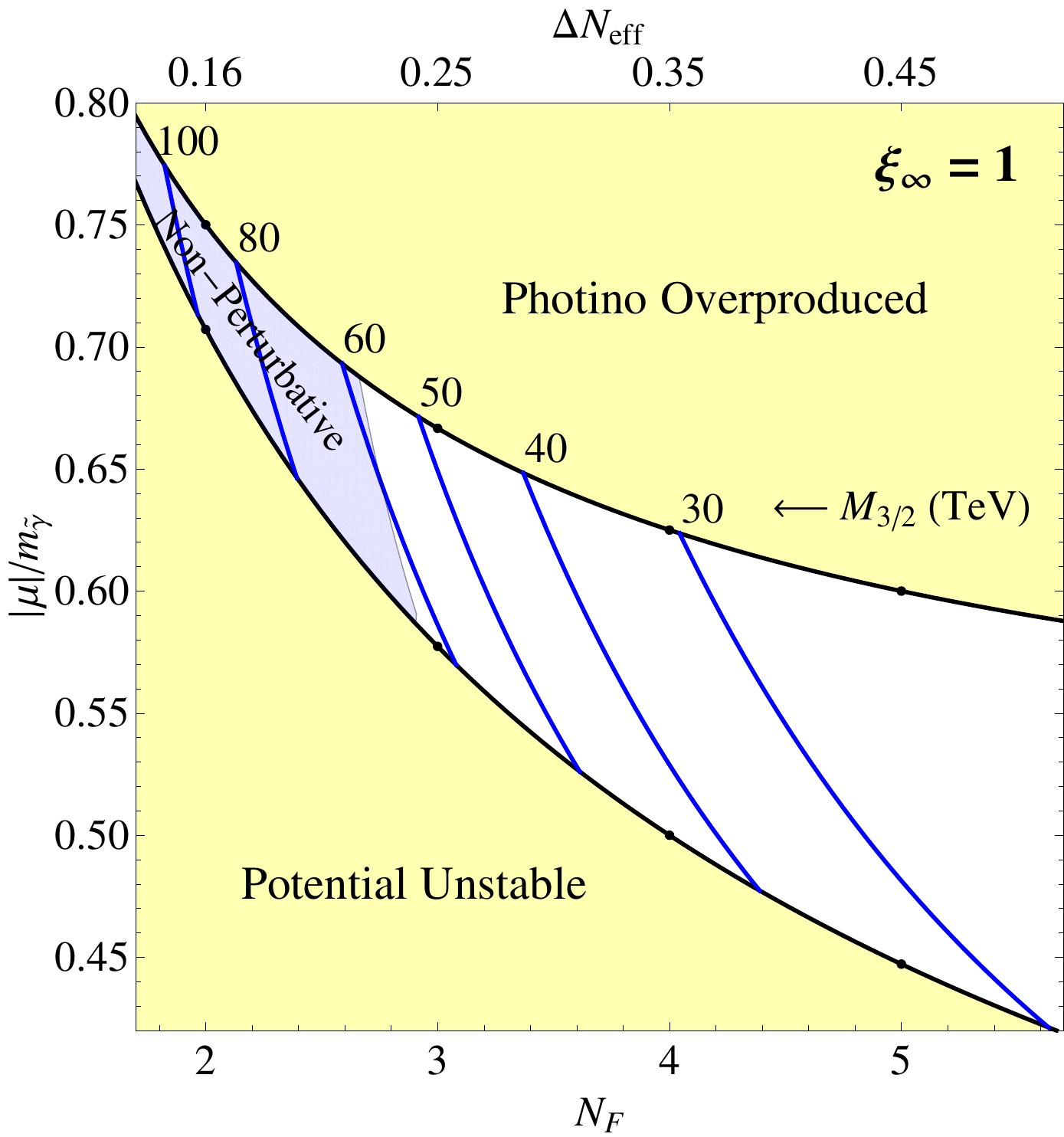}
\vspace*{-.1in}
\caption{Contours of constant $\mgravitino$ in the $(N_F,
  |\mu|/\mphotino)$ plane, with fixed $\xi_{\infty} = 1$, $\Omegatot
  \simeq 0.23$, and $g_*^{\text{v}}(T_f^{\text{v}}) =
  g_*^{\text{SM}}$.  $\Delta \neff$ is the effective number of extra
  neutrinos, as given by~\eqref{dneff}.  The shaded region labeled
  ``Non-Perturbative'' is forbidden by the considerations of
  self-interactions and perturbativity discussed
  in \secref{selfinteractions}.  For fixed $\xi_{\infty}$ and
  $\Omegatot$, $\mgravitino \propto
  g_*^{\text{v}}(T_f^{\text{v}})^{-1/6}$.
\label{fig:mgravitinoxi1}}
\end{figure}

In \figref{mgravitinoxi1}, we assume $\xi_{\infty} = 1$, and use
\eqref{xif} to determine $\xi_f$.  Requiring $\Omegatot \simeq 0.23$
fixes $\mgravitino$, and we present contours of constant $\mgravitino$
in the $(N_F, |\mu|/\mphotino)$ plane.  For large $N_F$, $\mgravitino$
becomes small, and requiring $\mgravitino \agt 34~\tev$, its lower
bound in minimal AMSB, implies $N_F \alt 5$.

These conclusions require significant assumptions, namely, that the
visible and hidden sectors were initially in thermal contact and then
decoupled, and that there is a desert. It is significant, though, that
even with these restrictive assumptions, there is a viable parameter
space with the correct dark matter relic density.  In addition, as we
will see in the next section, this simple scenario is especially
interesting, as it restricts the parameter space to regions of low
$N_F$ where the hidden sector may contribute significantly to extra
relativistic degrees of freedom, a prediction that will be tested in
the near future.

\section{Contributions to the Number of Relativistic Degrees of Freedom}
\label{sec:gstar}

The massless photons of the hidden sector contribute to the number of
relativistic degrees of freedom at any temperature.  Their existence
is therefore bounded by the standard theory of big bang
nucleosynthesis (BBN) and observations of the cosmic microwave
background (CMB), with testable consequences for future observations.

Constraints on the number of relativistic degrees of freedom are
typically reported as constraints on the effective number of extra
neutrinos
\begin{equation}
\Delta \neff \equiv \neff - \neff^{\text{SM}} \ ,
\end{equation}
where $\neff^{\text{SM}} \simeq 3.046$ deviates from 3 slightly
because the three neutrinos of the standard model are not completely
decoupled at the time of $e^+ e^-$ annihilation~\cite{Mangano:2005cc}.
Current bounds are
\begin{eqnarray}
\Delta \neff &=& 0.19 \pm 1.2 \ \text{(95\% CL) BBN} \\
\Delta \neff &=& 1.29^{+0.86}_{-0.88} \ \text{(68\% CL) CMB} \ ,
\end{eqnarray}
where the BBN constraint assumes a baryon density that has been fixed
to the value determined by the CMB, and both $^4$He and D data are
included~\cite{Cyburt:2004yc,Fields:2006ga}, and the CMB constraint is
the WMAP 7-year result~\cite{Komatsu:2010fb} that combines CMB
observations with distance information from baryon acoustic
oscillations, supernovae (SNIa) and Hubble constant measurements.  The
BBN and CMB results should be consistent, provided the number of
relativistic degrees of freedom remains constant between visible
sector temperatures of $T_{\text{BBN}} \sim \mev$ and $T_{\text{CMB}}
\sim \ev$.  The BBN result is fully consistent with the standard
model, while the CMB result shows a 1.5$\sigma$ excess.  In the near
future, Planck is expected to determine $\neff$ with a standard
deviation of $\sigma(\neff) \approx
0.3$~\cite{Hamann:2007sb,Ichikawa:2008pz,Colombo:2008ta,Joudaki:2011nw},
given only $\sim 1$ year of data.  This will, of course, be improved
with more data, and a future LSST-like survey may determine $\neff$
with a standard deviation of $\sigma(\neff) \approx
0.1$~\cite{Joudaki:2011nw}.  Needless to say, if the current WMAP
central value remains and the uncertainty shrinks as expected, these
measurements will have far-reaching consequences.

In hidden sector scenarios, $\Delta \neff$ is related to the number of
light hidden degrees of freedom by
\begin{equation}
\Delta \neff \ \frac{7}{8} \ 2 \ T_{\nu}^4 = g_*^{\text{h}}
(T_{\text{CMB}}^{\text{h}}) T_{\text{CMB}}^{\text{h}\, 4} \ ,
\end{equation}
where $T_{\nu} = (4/11)^{1/3} T_{\text{CMB}}^{\text{v}}$.  The
effective number of extra neutrinos is therefore
\begin{eqnarray}
\Delta \neff &=& \frac{4}{7} \left( \frac{11}{4} \right)^{\frac{4}{3}}
g_*^{\text{h}} (T_{\text{CMB}}^{\text{h}}) \,
\xi_{\text{CMB}}^4 \\
&=& \frac{4}{7} \left( \frac{11}{4} \right)^{\frac{4}{3}}
g_*^{\text{h}} (T_{\text{CMB}}^{\text{h}}) \,
\left[ \frac{g_*^{\text{h}}(T_f^{\text{h}})}
{g_*^{\text{h}}(T_{\text{CMB}}^{\text{h}})}
\frac{g_*^{\text{v}}(T_{\text{CMB}}^{\text{v}})}
{g_*^{\text{v}}(T_f^{\text{v}})} \right]^{\frac{4}{3}} \xi_f^4 \ .
\label{Nnu}
\end{eqnarray}
At the time of CMB decoupling,
$g_*^{\text{v}}(T_{\text{CMB}}^{\text{v}}) =
g_*^{\text{h}}(T_{\text{CMB}}^{\text{h}}) = 2$.  At freeze out, as
discussed in \secref{relicdensity}, $g_*^{\text{h}}(T_f^{\text{h}}) =
2$, but $g_*^{\text{h}}(T_f^{\text{v}})$ is more model-dependent.
Normalizing to $g_*^{\text{v}}(T_f^{\text{v}}) = g_*^{\text{SM}}$, we
find
\begin{equation}
\Delta \neff = \left(\frac{\xi_f}{2.60} \right)^4
\left[ \frac{106.75}
{g_*^{\text{v}}(T_f^{\text{v}})} \right]^{\frac{4}{3}} \ .
\end{equation}

\begin{figure}[tb]
\includegraphics[width=0.49\columnwidth]{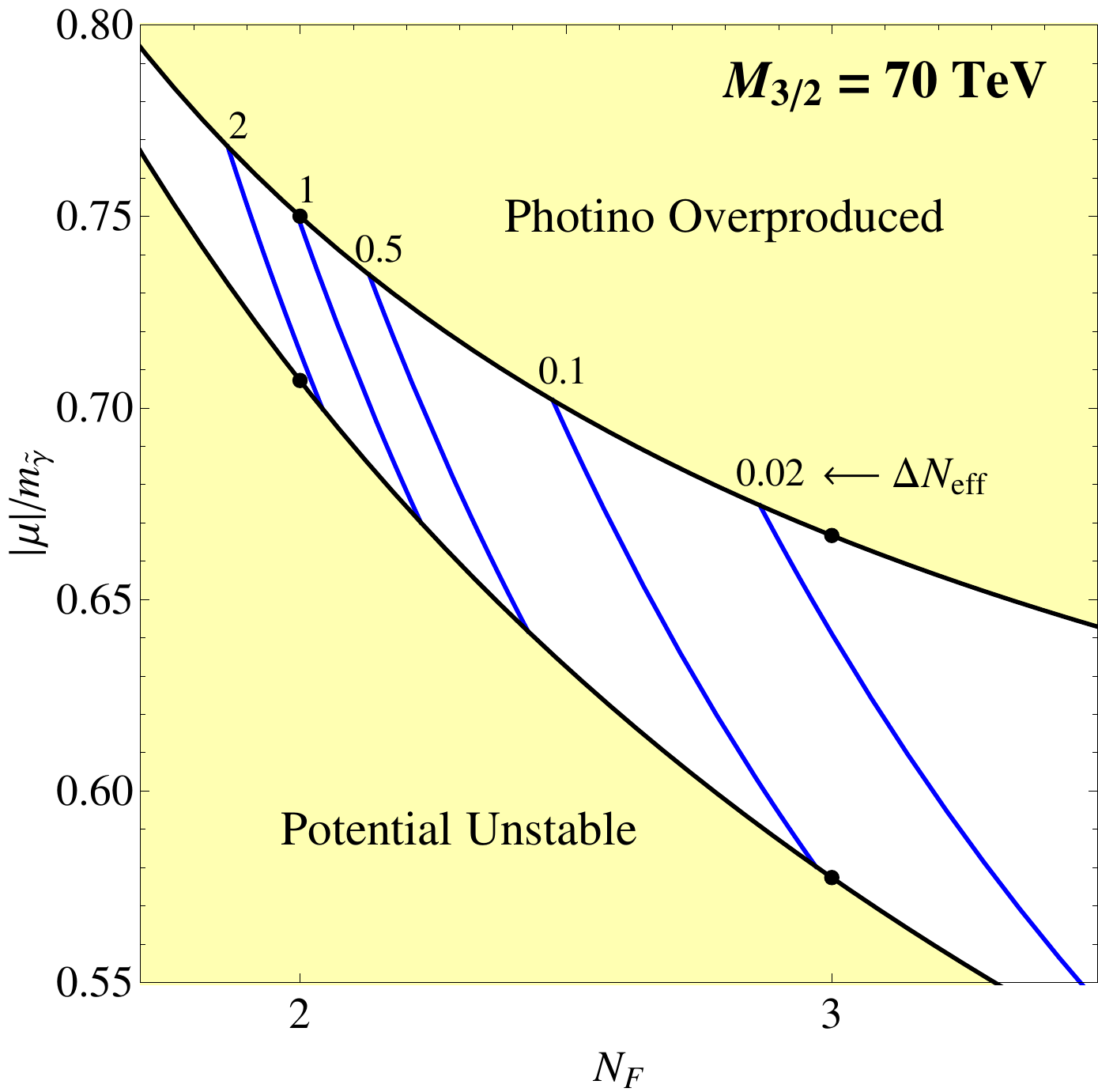}
\includegraphics[width=0.49\columnwidth]{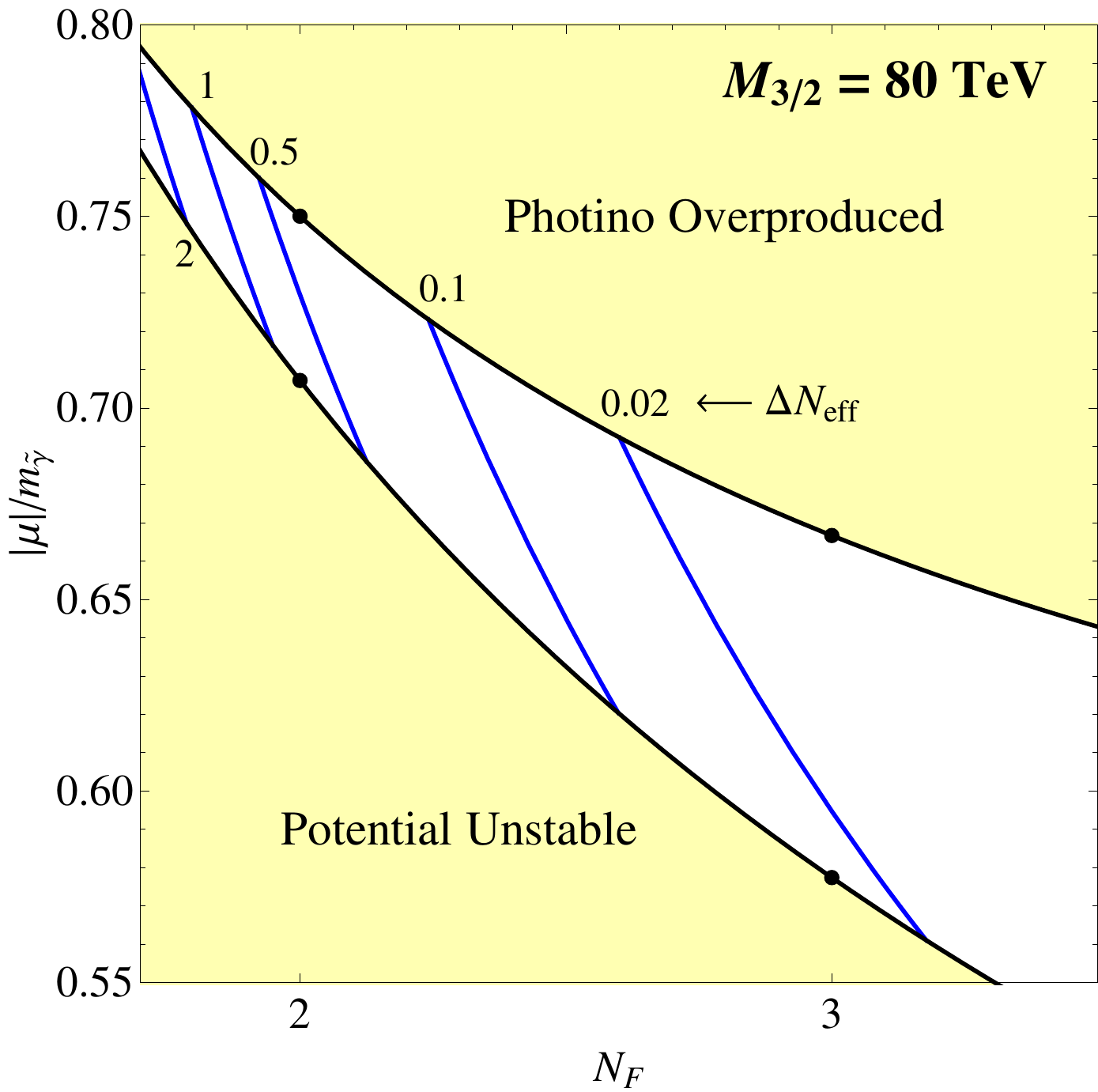}
\vspace*{-.1in}
\caption{Contours of constant $\Delta \neff$, the effective
number of extra neutrinos in the $(N_F, |\mu|/\mphotino)$ parameter space
for $\mgravitino = 70$ and 80 TeV, $\Omegatot \simeq 0.23$, and
$g_*^{\text{v}}(T_f^{\text{v}}) = g_*^{\text{SM}}$.  For fixed
$\Omegatot$, $\Delta \neff \propto \mgravitino^{-8} \,
g_*^{\text{v}}(T_f^{\text{v}})^{-4/3}$.
\label{fig:Nneutrinos}}
\end{figure}

Contours of $\Delta \neff$ are plotted in \figref{Nneutrinos} for
$\mgravitino = 70~\tev$ and 80 TeV.  Note that $\Delta \neff$ is
highly sensitive to $\mgravitino$; requiring a fixed $\Omegatot$,
$\Delta \neff \propto \mgravitino^{-8}$.  For the simplest case of
$N_F = 2$, however, we find that $\neff$ may have large deviations
from the standard model, with values up to $\Delta \neff \sim 2$
possible.  Much smaller values consistent with the standard model
within experimental uncertainties are also possible, however.

Finally, we note that the effective number of extra neutrinos may also
be expressed in terms of $\xi_{\infty}$.  Assuming a desert between
temperatures at very early times and the mass scales of the MSSM and
hidden SQED particles,
\begin{equation}
\Delta \neff = 0.20\left( \frac{N_F + \frac{8}{15}}{3}
\right)^{\frac{4}{3}} \xi_{\infty}^4 \ .
\label{dneff}
\end{equation}
This formula establishes a definite relation between $\Delta \neff$
and $N_F$, once $\xi_\infty$ has been fixed.  Such a prediction for
$\Delta \neff$ is independent of many aspects of the model, such as
$g_*^{\text{v}}(T_f^{\text{v}})$.  If a nonzero value for $\Delta
\neff$ is established, it will provide a strong motivation for hidden
sectors, and, in the current context, assuming $\xi_{\infty} = 1$, it
may be used to determine $N_F$.  Since fixing $\xi_\infty$ also
determines $\mgravitino$ unambiguously for every point on the
$(|\mu|/m_{\tilde\gamma},N_F)$ plane, collider bounds on $\mgravitino$
imply that $\Delta \neff$ is constrained too.  {}From
\figref{mgravitinoxi1} we can see that $\Delta \neff$ must be below
$\sim 0.4$ if indeed $\xi_\infty=1$ and if the dark matter content of
the Universe is to be explained by this model.  In the next section we
show that astrophysical constraints from self-interactions further
limit the parameter space, placing a lower bound on $\Delta \neff$ of
$\sim 0.2$, increasing the predictivity of $\Delta \neff$.

\section{Self-Interactions}
\label{sec:selfinteractions}

In calculating the relic density, the dark matter's coupling
$\alpha_X$ and mass $m_X$ drop out.  However, the hidden leptons and
sleptons are self-interacting through Coulomb interactions, and the
requirement that these self-interactions not violate observational
bounds imposes constraints on $\alpha_X$ and $m_X$.

Dark matter self-interactions are constrained by several observations,
and the case of long-range Coulomb interactions has been considered
recently in Refs.~\cite{Ackerman:2008gi,Feng:2009mn}.  The most
stringent constraint is from the observation of elliptical halos; dark
matter self-interactions cannot be so strong that the resulting energy
transfer between dark matter particles would have caused these halos
to relax to become spherical~\cite{2002ApJ...564...60M,Feng:2009mn}.

Neglecting sub-dominant $s$-channel contributions to the energy
transfer cross section, the relaxation time is~\cite{Feng:2009mn}
\begin{equation}
\tau_{\text{r}} \simeq \frac{m_X^3 v_0^3}{4 \sqrt{\pi} \alpha_X^2
  \rho_X C} \ ,
\end{equation}
where
\begin{equation}
C \approx \ln \left[
\frac{\left( b_{\text{max}} m_X v_0^2 \alpha_X^{-1} \right)^2 + 1}
{2} \right]
\end{equation}
is the Coulomb logarithm,
\begin{equation}
b_{\text{max}} \sim \frac{m_X v_0}{\sqrt{ 4 \pi \alpha_X \rho_X}}
\end{equation}
is the Debye screening length, and $v_0$ and $\rho_X$ are the velocity
dispersion and mass density, respectively, of dark matter particles at
a distance from the galactic center where the halo has been
established to be elliptical.  Following the analysis of
Ref.~\cite{Feng:2009mn}, observations of the elliptical galaxy NGC
720~\cite{Buote:2002wd,Humphrey:2006rv} have established an elliptical
halo at distances $r \sim 3-10~\kpc$ from the center, with mass
density $\rho_X(r) = 3.5 - 0.7~\gev/\cm^3$ and velocity $v_0 = 270 -
250~\km/\s$ in this range of radii.  Over this range, the estimate for
$\tau_{\text{r}}$ varies by a factor of 4.  Taking the values of
$\rho_X$ and $v_0$ at $r \sim 3~\kpc$, which yields the most stringent
limit, we find
\begin{equation}
\tau_{\text{r}} \simeq 9.0 \times 10^{9}~\yr
\left( \frac{m_X}{\tev} \right)^3
\left( \frac{0.01}{\alpha_X} \right)^2 \ \frac{90}{C} \ ,
\label{taur}
\end{equation}
where $C$ has been normalized to a typical value.  We may derive
bounds by requiring $\tau_{\text{r}} > 10^{10}~\yr$.

In our multi-component model of dark matter, the mass $m_X$ in
\eqref{taur} is ambiguous.  However, for nearly degenerate $m_{e_i}$
and $m_{\tilde{e}_i}$, we may take either one, of course, and for
highly non-degenerate cases, we may take the heavier mass $m_{e_i}$,
since the thermal relic density scales as $m_X^2$, and so the relic
density of the lighter species becomes insignificant.  For simplicity,
and given the other uncertainties of the analysis, we therefore
identify the lepton mass as the representative mass and set $m_X =
m_{e_i}$.

For fixed $\mgravitino$ and a given point in the $(N_F, |\mu| /
\mphotino)$ plane, $\alpha_X \propto m_{e_i}$.  Substituting this
relation in \eqref{taur}, requiring $\tau_{\text{r}} > 10^{10}~\yr$,
and setting the Coulomb logarithm to be $C = 90$, we find that the
observation of elliptical halos implies the lower bound
\begin{equation}
m_{e_i} > m_{\text{DM}}^{\text{min}} \equiv
\left[ \frac{6.6}{N_F \left( \mu / \mphotino \right)} \right]^2
\left( \frac{100~\tev}{\mgravitino} \right)^2 \tev \ .
\end{equation}

In addition, implicit in the formulae for the AMSB soft masses is the
assumption that the hidden sector particles discussed are in fact
present in the low-energy theory.  This requires that all superpartner
masses, in particular, the photino, be below $\mgravitino$, implying
$2 N_F \alpha_X / (4 \pi) < 1$.  This same criterion may be taken as
required to ensure perturbativity.  This requirement of perturbativity
is, again setting $C = 90$,
\begin{equation}
\frac{\mu}{\mphotino} > \left( \frac{0.66}{N_F} \right)^{\frac{2}{3}}
\frac{100~\tev}{\mgravitino} \ .
\label{perturbativity}
\end{equation}
If, for a certain region of parameter space, this upper bound on
$\alpha_X$ conflicts with the lower bound on $\alpha_X$ from halo
shapes, that region of parameter space is excluded.

\begin{figure}[tb]
\includegraphics[width=0.49\columnwidth]{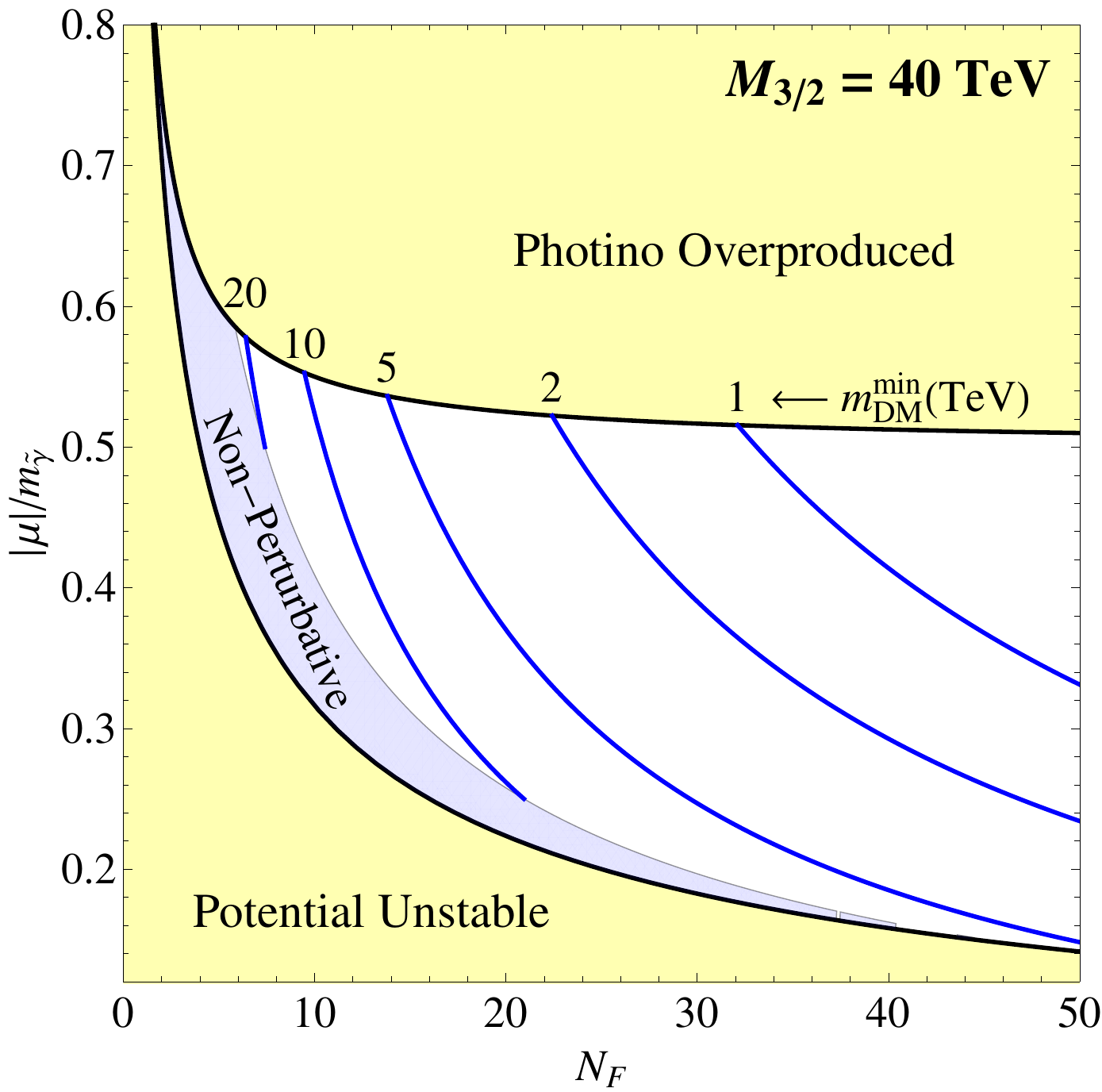}
\includegraphics[width=0.49\columnwidth]{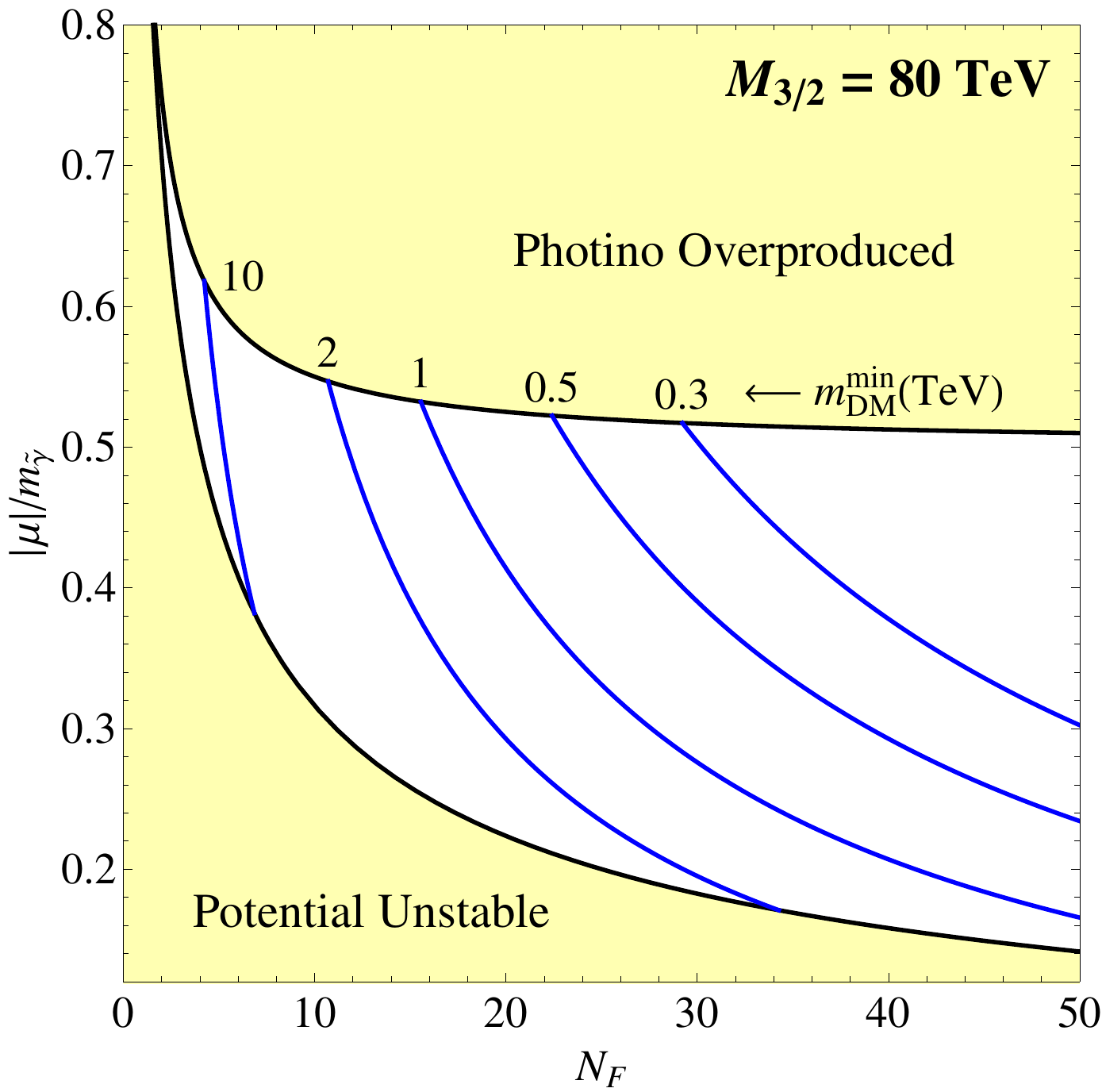} \\
\includegraphics[width=0.49\columnwidth]{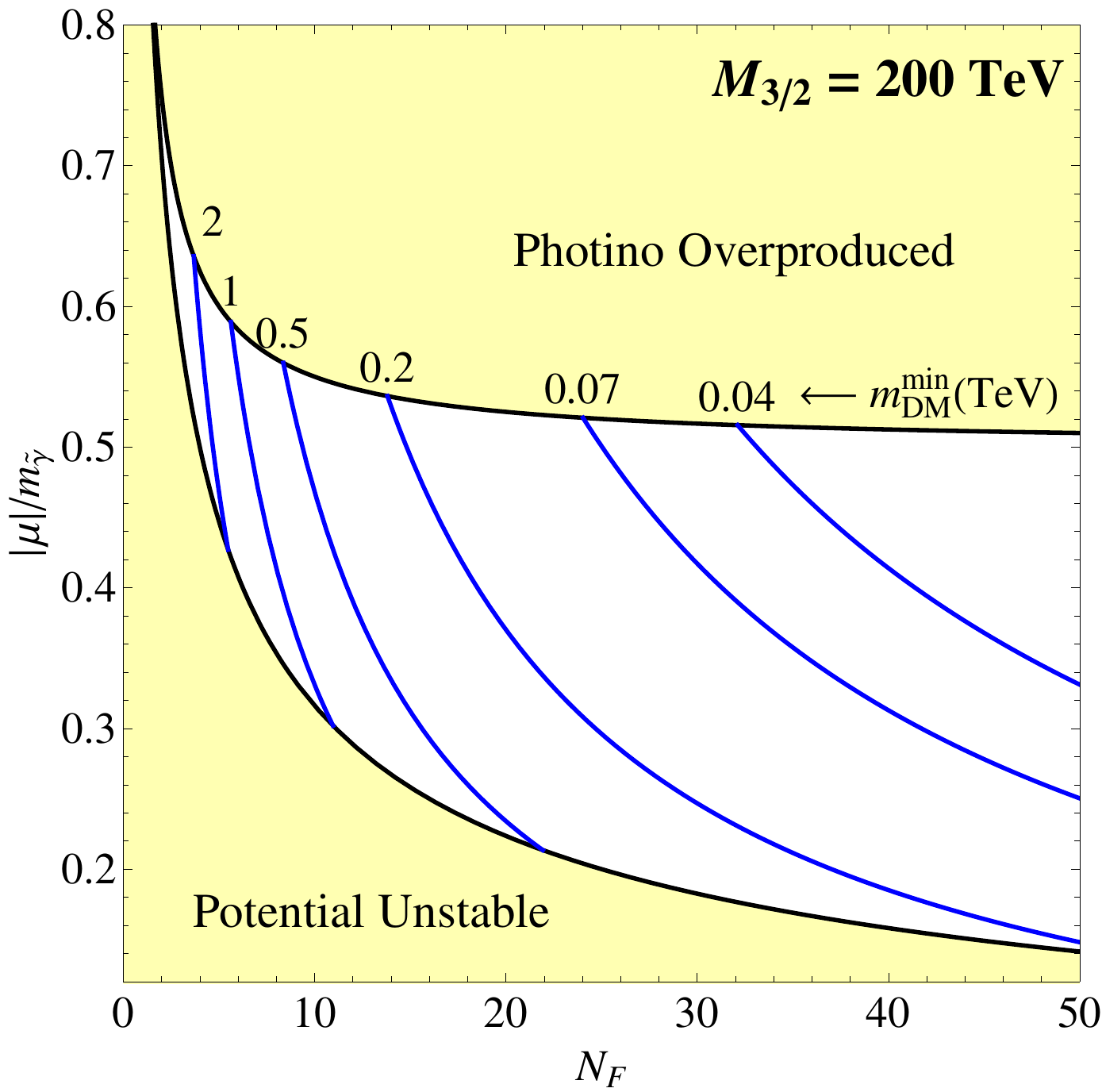}
\includegraphics[width=0.49\columnwidth]{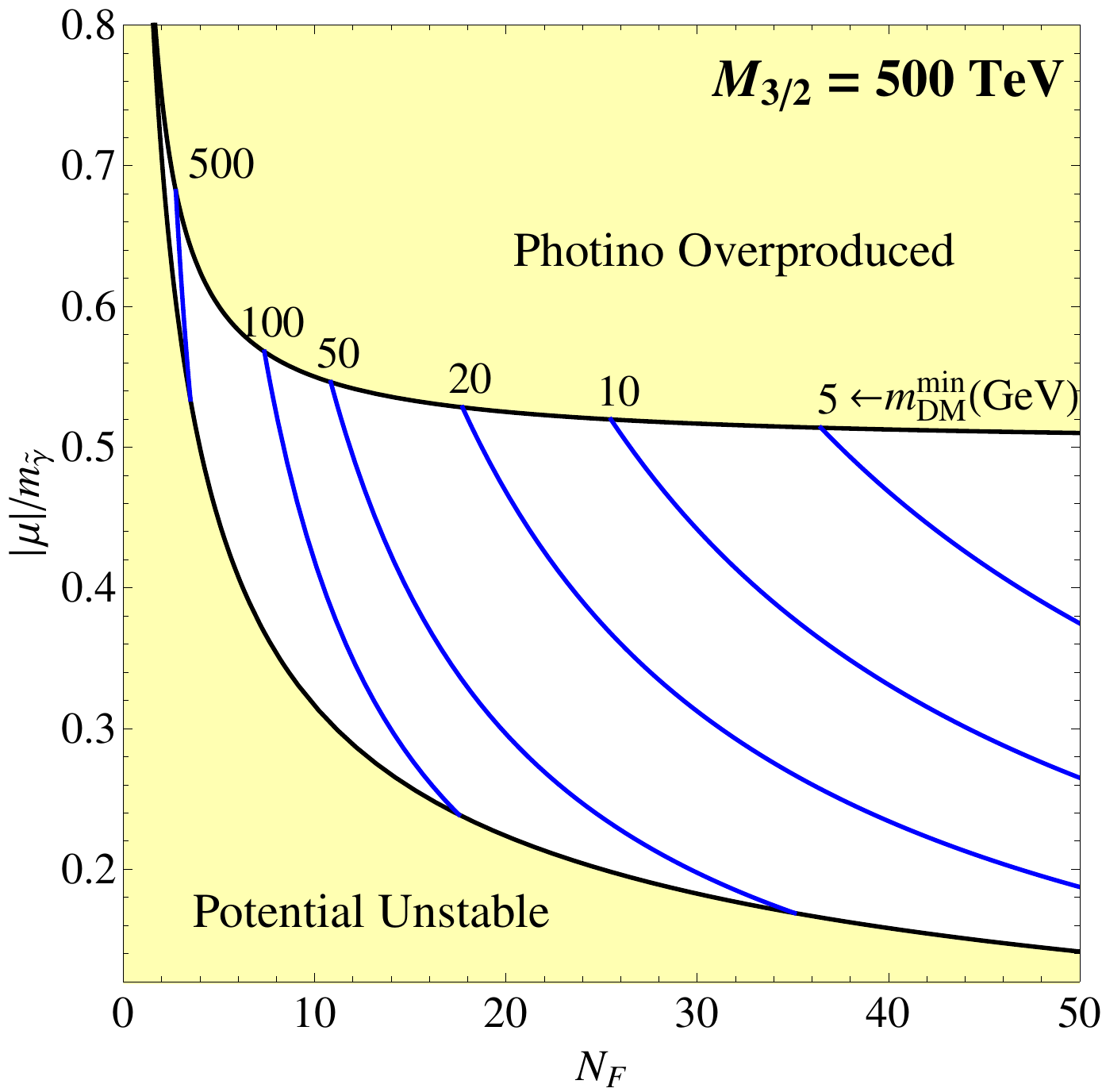}
\vspace*{-.1in}
\caption{Contours of constant $m_{\text{DM}}^{\text{min}}$, the
minimal lepton mass that is consistent with self-interaction bounds
from the observation of elliptical halos, for $\mgravitino = 40, 80,
200$, and 500 TeV, and $\Omegatot = 0.23$.  In the shaded region
marked ``Non-Perturbative,'' $\mphotino > \mgravitino$, signaling a
breakdown of perturbativity. Note that the values of
$m_{\text{DM}}^{\text{min}}$ are in units of GeV for the $\mgravitino
= 500~\tev$ panel, and in TeV for the others, as indicated.
\label{fig:mDMmin}}
\end{figure}

Lower bounds on the dark matter mass and a region of
non-perturbativity are shown in \figref{mDMmin}.  {}From the
$\mgravitino = 40~\tev$ panel, we see that the perturbativity
constraint excludes otherwise viable regions in the parameter space
with low $N_F$ if $\mgravitino$ is very low.  We find, however, that
it does not constrain the allowed parameter space with $N_F \ge 2$ if
$\mgravitino\agt 65~\tev$.  {}From the $\mgravitino = 80~\tev$ panel,
we see that the regions with large $\Delta \neff \sim 1$ of
\figref{Nneutrinos} have heavy dark matter, with
$m_{\text{DM}}^{\text{min}} \sim 8 - 30~\tev$.

Finally, from all the panels, we see that as $N_F$ increases,
$m_{\text{DM}}^{\text{min}}$ decreases.  For a fixed $|\mu| /
\mphotino$, increasing $N_F$ implies lower $\alpha_X$, weakening the
self-interaction constraint, and making lighter dark matter masses
possible.  {}From the $\mgravitino = 200$ and 500 TeV panels, we see
that dark matter masses as low as $\sim 10~\gev$ are possible.  For
$\mgravitino \sim 500~\tev$, light dark matter is allowed for $N_F
\sim 30$ by all constraints.  In contrast to WIMPs, which have masses
typically in the range $100~\gev - 1~\tev$, this WIMPless scenario
allows a much broader range of dark matter masses, while
preserving the required relic density.  The lower bound is imposed by
the requirement that the self-interactions of charged dark matter not
be too large, which is, of course, absent for WIMPs.  Even so, the
possibility of $10~\gev$ WIMPless dark matter remains, and is
particularly interesting, given current hints of signals in this range
from DAMA/LIBRA~\cite{Bernabei:2008yi,Bernabei:2010mq} and
CoGeNT~\cite{Aalseth:2010vx,Aalseth:2011wp}.  It would be interesting
to investigate the possibility of adding connector fields charged
under both visible sector gauge groups and the hidden U$(1)$
to explain these signals.

At this point it is worth revisiting the discussion in \secref{gstar}
regarding $\Delta \neff$.  As mentioned before, once $\xi_\infty$ is
fixed, $\mgravitino$ is determined for every point on the
$(|\mu|/m_{\tilde\gamma},N_F)$ plane.  We may then use
\eqref{perturbativity} to check whether such a point satisfies the
requirement of perturbativity.  The region excluded by this
requirement is indicated in \figref{mgravitinoxi1}.  {}From this we
conclude that if $\xi_\infty=1$, the model predicts that $\Delta
\neff$ lies in the range $0.2 \sim 0.4$.  Such a prediction will
definitely be tested in the near future, as discussed in
\secref{gstar}.

\section{Discussion}
\label{sec:discussion}

In this work, we have constructed an extremely simple model of dark
matter in AMSB with a hidden sector.  The hidden sector is SQED with
$N_F$ degenerate flavors of leptons and sleptons. Through a
combination of electric charge, lepton flavor, and $R$-parity
conservation, all $N_F$ flavors of leptons and sleptons are stable and
contribute to cold dark matter.  As in the case of WIMPs, the thermal
relic density of these particles is directly related to the mechanism
of electroweak symmetry breaking and is naturally of the right order
of magnitude to be dark matter.  In contrast to WIMPs, however, the
dark matter has a mass that may be anywhere from $\sim 10~\gev$ to 10
TeV, and it also has interesting astrophysical implications that are
absent for WIMPs.  In particular, it self-interacts through Coulomb
interactions, and the massless hidden photon implies extra
relativistic degrees of freedom in the range $ \Delta \neff \sim 0 -
2$, with significant deviations from 0 possible in the simplest models
with low $N_F = 2 - 5$.  If the hidden and visible sectors were in
thermal contact in the very early Universe, $\Delta \neff$ is more
constrained and is predicted to assume values in the range $0.2 -
0.4$.  Such predictions are especially interesting, given that data
already taken by Planck may provide evidence of such deviations.

As dark matter in this model consists of hidden particles, the visible
sector is relieved of the duty of providing dark matter.  This is a
welcome possibility, as the most natural dark matter candidate in
AMSB, the neutral Wino, has a thermal relic density that is typically
too small to be all of dark matter by several orders of magnitude.
$R$-parity may also be broken in the visible sector, provided
constraints from proton decay and elsewhere are not violated, with
implications for collider searches.

In this work, we have introduced $\mu$, a supersymmetric mass for the
hidden sleptons, and treated it as a free parameter.  This is a
time-honored approach and has yielded interesting results.  That said,
just as a complete neutralino WIMP scenario requires a solution to the
$\mu$ problem in the visible sector, a complete WIMPless scenario
requires a concrete $\mu$-term mechanism in the hidden sector.
Formally, to realize the WIMPless miracle with a relic density that is
independent of $g$, the $\mu$ parameter must have the form $\mu \sim
\mphotino \sim (g^2 / 16 \pi^2) \mgravitino$. Of course, we have found
that constraints from dark matter self-interactions require $\mphotino
\sim 10~\gev - 10~\tev$, and so a $\mu$-term of the form $\mu \sim (1
/ 16 \pi^2) \mgravitino$, while parametrically incorrect, will yield
phenomenologically viable numerical values for $\mu$. There are
several proposed mechanisms for generating $\mu$-terms in the visible
sector of AMSB models; see, for example,
Refs.~\cite{Pomarol:1999ie,Chacko:1999am,Katz:1999uw}.  In these
mechanisms, the form of $\mu$ is typically $f(\lambda_i)/ (16 \pi^2)
\mgravitino$, where $f(\lambda_i)$ is a function of new Yukawa
couplings.  For $f(\lambda_i) \sim 1$, these mechanisms solve the
$\mu$ problem in the MSSM, and similar mechanisms in the hidden sector
would provide numerically correct $\mu$-terms for this model of
WIMPless dark matter.  It would be more satisfying, however, to find
$\mu$-term mechanisms that provide the parametrically correct form
$\mu \sim (g^2 / 16 \pi^2) \mgravitino$.  This is beyond the scope of
this work, but we note that, in contrast to the traditional $\mu$
problem of the visible sector, which requires both $\mu \sim \mweak$
and $\sqrt{B} \sim \mweak$, in the hidden sector, the constraints on
$B$ are less stringent, which may open new avenues for $\mu$-term
generation.

Finally, we note that we have made several assumptions in the interest
of simplicity. More general models with non-universal charges $q_i$,
non-universal $\mu$-terms, or non-vanishing $B$-terms may have
interesting features.  In addition, we have assumed that there are no
interactions between the visible and hidden sector, which eliminates
direct and indirect dark matter signals.  One could introduce
connector particles with both visible and hidden sector charges to
induce such signals.  A special motivation for introducing such
interactions is that, as shown in \secref{selfinteractions}, this
framework provides the possibility of dark matter particles with
a naturally correct thermal relic density, but masses $\sim
10~\gev$, that is, at the scale indicated by current direct detection
anomalies.  Such connector particles would also induce mixing between
the hidden and visible photons, with a host of possible implications.

\section*{Acknowledgments}

We thank Manoj Kaplinghat, Yuri Shirman, and especially Yael Shadmi
for many helpful conversations.  The work of JLF was supported in part
by NSF grants PHY--0653656 and PHY--0970173. The work of VR was
supported in part by DOE grant DE-FG02-04ER-41298. The work of ZS was
supported in part by DOE grant FG-03-94ER40837.

\providecommand{\href}[2]{#2}\begingroup\raggedright\endgroup

\end{document}